\documentclass{ws-p8-50x6-00}
\def\lapproxeq{\lower .7ex\hbox{$\;\stackrel{\textstyle
<}{\sim}\;$}}
\def\gapproxeq{\lower .7ex\hbox{$\;\stackrel{\textstyle
>}{\sim}\;$}}

\begin{document}

\title{Structure of the final states in Higgs production at Hadronic Colliders}

\author{V.A. Khoze}

\address{Department of Physics and Institute for Particle Physics Phenomenology,
University of Durham, Durham, DH1 3LE, U.K.\\
E-mail: V.A.Khoze@durham.ac.uk}


\maketitle

\abstracts{ We quantify the rate and the signal-to-background
ratio for Higgs $\rightarrow b\bar{b}$ detection in
double-diffractive events at the Tevatron and the LHC.  The signal
is predicted to be very small at the Tevatron, but observable at
the LHC.  We exemplify the diagnostic power of hadronic antenna
pattern by considering the topology of hadronic flow corresponding
to Higgs $\rightarrow b\bar{b}$ events.}

One of the biggest challenges facing the high-energy experiments
is to find a good signal with which to identify the Higgs boson.
Following the closure of LEP2, the focus of searches for the Higgs
is concentrated on the measurements at the present and forthcoming
hadron colliders, the Tevatron and the LHC.

To ascertain whether a Higgs signal can be seen, it is crucial to
show first that the background does not overwhelm the signal.  For
instance, as well known, an observation of the inclusive
intermediate mass Higgs production, that is $pp$ or $p\bar{p}
\rightarrow HX$ with $H \rightarrow b\bar{b}$ is considered to be
impossible because of an extremely small signal-to-background
ratio due to gluon-gluon fusion, $gg \rightarrow b\bar{b}$.  One
possibility is to observe the Higgs in association with massive
particles ($W/Z$, $t$-quarks).  Here we discuss two (less
conventional) approaches to tackle the background problem in the
case of central $H \rightarrow b\bar{b}$ production.

One way to reduce QCD background, is to study the central
production of the Higgs in events with a large rapidity gap on
either side, see \cite{KMR1} and references therein.  An obvious
advantage of the rapidity gap approach is the spectacularly clean
experimental signatures:  hadron-free (`no-flight') zones between
the remnants of the incoming protons and the produced system.

Another possibility is to use the topology of hadronic flows
corresponding to Higgs production to dissect the colour structure
of the final state \cite{HKS}.  Topologometry of hadronic flows
(event portrait) proves to be a very successful discriminative
tool to probe the detailed properties of multi-jet events in
various hard processes, see \cite{VAK} and references therein.

We begin with double-diffractive processes of the type
\begin{equation}
\label{eq:a1}
 pp \; \rightarrow \; p \: + \: M \: + \: p,
\end{equation}
where the protons remain intact.  They allow the reconstruction of
the ``missing'' mass $M$ (say a Higgs boson) with good resolution,
and so provide an ideal way to search for new resonances and
threshold behaviour phenomena \cite{KMR2}.  Moreover, in exclusive
processes with forward protons the incoming $gg$ state satisfies
special selection rules, namely it has $J_z = 0$, and positive $C$
and $P$ parity.  These selection rules are of crucial importance
in suppressing the QCD background when searching for the $H
\rightarrow b\bar{b}$ signal.

In order to use the `missing-mass' method to search for a Higgs
boson, via the $H \rightarrow b\bar{b}$ decay mode, we have to
estimate the QCD background which arises from the production of a
pair of jets with invariant mass about $M_H$.  The good news is
that the signal-to-background ratio does not depend on the
uncertainty in the soft rescattering effects, and is given just by
the ratio of the corresponding $gg \rightarrow H \rightarrow
b\bar{b}$ and $gg \rightarrow b\bar{b}$ subprocesses.

A remarkable advantage of the signature (\ref{eq:a1}) for the $H
\rightarrow b\bar{b}$ events is that here the $H \rightarrow
b\bar{b}$ signal/$b\bar{b}$ background ratio is strongly enhanced
due to colour factors, gluon polarization selection and the spin
$\frac{1}{2}$ nature of quarks.  An explicit calculation assuming
$M_H = 120$~GeV and imposing the $\theta > 60^\circ$ cut of low
$E_T$ jets, gives a signal-to-background ratio
\begin{equation}
\label{eq:a2}
 \frac{S}{B_{b\bar{b}}} \; \gapproxeq \; 4 \: \left ( \frac{1~{\rm GeV}}{\Delta
 M}\right ).
\end{equation}
The signal is, thus, in excess of background even at mass
resolution $\Delta M \sim 2$~GeV, so the $b\bar{b}$ background
should not be a problem.  Unfortunately the situation worsens for
inclusive Higgs production, where the polarization arguments
become redundant.  In this case $S/B_{b\bar{b}}$ ratio is
additionally suppressed by a factor $\sim 20-30$.

While the predictions for the $S/B_{b\bar{b}}$-ratio look quite
favourable for Higgs searches using the missing-mass method, the
expected event rate casts a shadow on the feasibility of this
approach (at least for experiments at the Tevatron).  In our
recent analysis we found
\begin{eqnarray}
\label{eq:a3}
 \sigma_H \; = \; \sigma (p\bar{p} \rightarrow p + H + \bar{p}) \;
 \simeq \; 0.06~{\rm fb} \quad\quad & {\rm at} & \quad\quad \sqrt{s} \; = \; 2~{\rm
 TeV}, \\
\label{eq:a4}
 \sigma_H \; = \; \sigma (pp \rightarrow p + H + p) \;
 \simeq \; 2.2~{\rm fb} \quad\quad & {\rm at} & \quad\quad \sqrt{s} \; = \; 14~{\rm
 TeV}
\end{eqnarray}
for a Higgs boson of mass 120~GeV.  These values correspond to the
cross section ratio
\begin{equation}
\label{eq:a5}
 \sigma_H/\sigma_{{\rm incl}\:(gg \rightarrow H)} \; \simeq \;
 10^{-4}
\end{equation}
and are much lower than the predictions to other authors.  However
the recent CDF study of diffractive dijet production \cite{CDF}
provides strong experimental evidence in favour of our pessimistic
estimates.

We emphasize that low expected signal cross section at the
Tevatron just illustrates the high price to be paid for improving
the $S/B_{b\bar{b}}$ ratio by selecting events where the protons
remain intact.  Nevertheless, there is a real chance to observe
double-diffractive Higgs production at the LHC, since both the
cross section and the luminosity are much larger than at the
Tevatron.  For an integrated luminosity of 100~fb$^{-1}$ at the
LHC we expect, for $M_H = 120$~GeV, about 200 $H \rightarrow
b\bar{b}$ events.

Let us now turn to hadronic radiation pattern for $H \rightarrow
b\bar{b}$ production at hadronic colliders.  This can be used as a
discriminative tool to distinguish between the signal and the
conventional QCD background.  In \cite{HKS} the radiation samples
were compared for the signal $(gg \rightarrow H \rightarrow
b\bar{b} + g)$ and background $(gg \rightarrow b\bar{b} + g)$
production.  It was found that the main difference between these
two came from the radiation {\it between} the final-state jets. In
particular, there is approximately 4/3 more radiation between the
final-state jets for the Higgs production.  This is due to the
absence of a colour connection between the quarks in the
background process.

The radiation patterns were studied also for Higgs production in
association with a $W$ boson in the process
\begin{equation}
\label{eq:a6}
 q\bar{q}^\prime \; \rightarrow \; W^* \; \rightarrow \; W
 (\rightarrow \ell \nu_\ell) \: H (\rightarrow b\bar{b})
\end{equation}
and for the background process
\begin{equation}
\label{eq:a7}
 q\bar{q}^\prime \; \rightarrow \; W (\rightarrow \ell \nu_\ell)
 \: + \: b\bar{b},
\end{equation}
when $M_{b\bar{b}} \sim M_H$.  Again the signal and background
processes are shown to have quite different radiation patterns.

\section*{Acknowledgements}

It is a pleasure to thank the Organizing Committee of ISMD 2001
for the kind invitation and hospitality in Datong.  I am grateful
to The Leverhulme Trust for a Fellowship.


\end{document}